\documentclass{llncs}

\usepackage[T1]{fontenc}
\usepackage{graphicx}
\usepackage{multirow}
\usepackage{subcaption}
\usepackage{caption}
\usepackage{balance}
\usepackage{multicol}
\usepackage{makecell}
\usepackage{url}
\usepackage{geometry}
\usepackage{enumitem}

\usepackage{array}
\usepackage{tabularx}

\begin{document}
\title{THAPI: Tracing Heterogeneous APIs }

\author{{ Solomon Bekele}\inst{1} \and
       { {Aurelio Vivas}\inst{2}  \and
       { Thomas Applencourt}\inst{1}{} \and
       { Servesh Muralidharan}\inst{1} \and
       { Bryce Allen} \inst{1}\and
       { Kazutomo Yoshii}{inst{1}}
       { Swann Perarnau}\inst{1} \and
       { Brice Videau}\inst{1}}
    }
    
%Second Author\inst{2,3}\orcidID{1111-2222-3333-4444} \and
%Third Author\inst{3}\orcidID{2222--3333-4444-5555}}
%
%\authorrunning {}

%Springer Heidelberg, Tiergartenstr. 17, 69121 Heidelberg, Germany
             % typeset the header of the contribution
%
\institute{Argonne National Laboratory \\
    \email{\{sbekele,tapplencou,servesh,ballen,kazutomo@mcs,swann,bvideau\}.anl.gov} 
    \and
    University De Los Andes - Colombia \\
    \email{aa.vivas@uniandes.edu.co} }

      \authorrunning{S. Bekele et al.}

\maketitle 
\begin{abstract}
%The computational power of high-performance computing systems (HPCs) leaped a big stride over the last decade. However, this remarkable performance enhancement came entangled with increased system complexity. These systems now integrate heterogeneous computing components and diverse programming models. The programming models are usually implemented on top of each other, making performance analysis and debugging increasingly challenging. This paper proposes THAPI (Tracing Heterogeneous APIs), a portable, programming model-centric tracing framework that provides detailed programming model insights for debugging and performance optimization in heterogeneous HPC environments. THAPI traces comprehensive API call details across major programming models, ensuring fine-grained introspection and programming model context capture. The tracing framework is built on the Linux Trace Toolkit Next Generation (LTTng) for efficiency and scalability. THAPI also provides analysis tools that leverages Babeltrace2 to deliver a holistic view of execution, making it an essential tool for diagnosing performance bottlenecks and optimizing system behavior.

As we reach exascale, production High Performance Computing (HPC) systems are
increasing in complexity. These systems now comprise multiple heterogeneous
computing components (CPUs and GPUs) utilized through diverse, often
vendor-specific programming models. As application developers and programming
models experts develop higher-level, portable programming models for these
systems, debugging and performance optimization requires understanding how
multiple programming models stacked on top of each other interact with one another. This paper
discusses THAPI (Tracing Heterogeneous APIs), a portable, \textit{programming
model-centric} tracing framework: by capturing comprehensive API call details
across layers of the HPC software stack, THAPI enables fine-grained
understanding and analysis of how applications interact with programming models
and heterogeneous hardware. Leveraging state of the art tracing framework like the Linux Trace Toolkit Next Generation (LTTng) and tracing
much more than other tracing toolkits, focused on function names and timestamps,
this approach enables us to diagnose performance bottlenecks across the software
stack, optimize application behavior, and debug programming model implementation
issues.

\end{abstract}
\section{Introduction}

The pursuit of exascale computing and the broader evolution of high-performance computing has led to massive computational capabilities. With high performance and energy efficiency in mind, these systems are being designed with a mixture of CPUs, GPUs, FPGAs, and other accelerators sourced from different manufacturers such as Intel, NVIDIA, AMD and others. As of November 2024, nine of the top ten fastest systems in the TOP500 list are heterogeneous \cite{top500}. The heterogeneity at the hardware level  pushes the diversity in the programming environments, expanding the spectrum of programming models available for high-performance computing applications. These programming models layer upon one another, resulting in a tightly integrated system that adds to its intricacy. Portable models like Kokkos~\cite{kokkos} target Intel, NVIDIA, and AMD GPUs using SYCL~\cite{Sycl:2020}, CUDA~\cite{cuda:2008}, and HIP~\cite{hip} backends, respectively. There are portable implementations of programming models that extend support to architectures they were not originally designed for.  For example, HIP is designed for AMD GPUs, and now there are experimental implementations, such as HIPLZ~\cite{hipzl} and HIPCL~\cite{hipcl}, which extend its compatibility to Intel GPUs by leveraging Level-Zero and OpenCL, respectively. This growing complexity in programming environments makes the task of introspecting and analyzing the interactions among programming models -- and their relationship with applications -- more challenging. %\section{Background and Motivation}
\subsection{Background and Motivation}

 To analyze performance, debug errors, and resolve performance issues, it is important to understand the performance of applications across different programming models. This includes  identifying potential sources of inefficiencies that may arise from layering of APIs, runtime translations and architectural differences. 

Vendors offer tools specific to their products, such as Intel’s vTune \cite{WinNT}, Nvidia’s Nsight \cite{Nsight}, AMD's ROCprof \cite{rocprof}. Although these tools work well in their respective environments, they do not have the capability to work with portable applications on a variety of platforms and programming models.

Performance analysis tools, with cross-platform support help fill this gap. Tools like TAU \cite{tau}, HPCToolkit \cite{hpctoolkit}, and Score-P \cite{scorep} are third-party tools that provide performance profiling and tracing capabilities for HPC systems. These tools gather performance information through instrumentation and sampling, and provide insight through their analysis and visualization tools. While these tools offer robust performance analysis features, they capture only limited information about lower-level programming model context, which is essential for runtime developers and system programmers. Their primary focus is the timing of API calls rather than  the complete call context. 

\begin{footnotesize}
\begin{center}
\begin{minipage}{0.95\linewidth}
\begin{verbatim}
THAPI: 21:41:26.240059291 - x4204c0s1b0n0 - vpid: 124765, vtid: 124765
       - lttng_ust_ze:zeCommandListAppendMemoryCopy_entry: {hCommandL
       ist: 0x000000000508aea8, dstptr: 0xff007ffffff90000, srcptr: 
       x00007fffedceab98, size: 472, hSignalEvent: 0x0000000005165898,
       numWaitEvents: 0, phWaitEvents: 0x0000000000000000, phWaitEvent
       s_vals: [  ] } }
       
TAU: {"event-type": "entry", "name": "zeCommandListAppendMemoryCopy",
     "time": "4710005.000000", "node-id": "0", "thread-id": "2" }
\end{verbatim}
\end{minipage}
\end{center}
\end{footnotesize}

For instance, the  plain text above illustrates the trace event content for the entry of Level-Zero API call {\tt zeCommandListAppendMemCopy} as captured by both THAPI and TAU during the execution of the  505.lbm\_r benchmark from SPEChpc 2021. This example highlights the difference in the level of detail recorded by the two tools.  TAU captures minimal information in regard to the call, focusing on its metadata (name, timestamp, node-id, thread-id ). Whereas THAPI records the detailed API call information: detailed arguments: source and destinations pointers, transfer size, command list handle and metadata (timestamp, node-id, process-id, thread-id, name).   For instance, from these details, we can deduce that the operation is data transfer from host to device, as indicated by the memory addresses: the source pointer starts with 0x00, indicating host memory, while the destination pointer begins with 0xff, implying device memory. We also know the size of the transfer and more. These low-level details are essential for reconstructing the execution flow, ensure reproducibility,  and detecting errors and unexpected behavior. We will demonstrate later how such detailed information can be valuable. 

In this paper, we propose THAPI, a \textit{programming model-centric}, tracing framework for heterogeneous HPC systems. THAPI  supports the aforementioned variety of platforms — diverse hardware, heterogeneity, programming models and workloads —  and helps programmers and system designers to understand applications’ performance,  debug errors, identify performance bottlenecks, and find potential opportunities for optimization.  THAPI captures as much context as possible while maintaining minimal overhead.

As we mentioned earlier, tools like  TAU \cite{tau}, Score-P \cite{scorep} and HPCToolkit \cite{hpctoolkit}  provide  a wide range of functionalities, THAPI, however, complements these tools with the following contributions:

\begin{itemize}[label=\textbullet]
\item For Runtime Developers and Performance Engineers: THAPI provides a portable tracing framework that captures the low-level programming model context that is essential to understand runtime behavior of applications. It collects all API calls along with their arguments --input and output pointers, values behind pointers, etc --  facilitating the introspection of the interaction between layered programming models (For example,  HIP on top of Level-Zero backend for HIPLZ). Additionally, it profiles GPU execution, offering a comprehensive view of heterogeneous runtime behavior.

\item For Tools Developers: THAPI demonstrates automatic generation of tracepoints and analysis tools plugins from the programming model headers simplifying the instrumentation process and the maintenance of the tool. 

\item For Application Developers: Complementary analysis plugin tools generated automatically from the programming models that can produce portable summary and timeline visualization.

\item  Sampling framework that captures rich execution context by reading GPU performance counters.

\iffalse

Detailed Tracing Information: As highlighted earlier, THAPI records API calls with detailed arguments, such as kernel parameters, pointers, configuration details, and values behind the pointers, enabling accurate recreation of the programming model context. In contrast, state-of-the-art tools that we explored focus on the timing of the API calls for the purpose of profiling (to analyze where the execution times spent). However, they do not log detailed API arguments, making THAPI particularly effective for debugging.

\item Instrumentation and Integration: THAPI eliminates the need for source code instrumentation. Tracepoints are automatically generated from programming model header files, minimizing developer effort. This approach contrasts with the aforementioned tools,  which require more intrusive integration. 

\fi

\end{itemize}

THAPI supports a wide range of heterogeneous programming models, including CUDA, OpenCL, HIP, Level-Zero as well as hybrid parallel programming models such as MPI and OpenMP. Its modular code architecture allows for seamless additions of future programming models and facilitates continuous enhancements.

The rest of the paper is organized as follows:   Section~\ref{sec:related} delves into prior works centered on performance analysis and debugging within HPC. Section~\ref{sec:thapi} discusses the details of our tracing framework. Section~\ref{sec:case} presents various test cases demonstrating the effectiveness of the tool. In Section~\ref{sec:eval},  we showcase the assessment results of our proposed framework,  and Section~\ref{sec:co} wraps up the paper with a conclusion.

\section{Related Work}\label{sec:related}

Historically,  performance analysis tools have focused on CPUs and have been more advanced for them. However, the emergence of heterogeneous systems has driven the development of tools targeting these architectures. There has been a body of work on performance tools for heterogeneous systems. Vendors like NVIDIA, Intel, and AMD offer several tools that provide monitoring capabilities targeting their GPUs, CPUs, or both. NVIDIA's Nsight~\cite{Nsight} and Nsight System~\cite{Nsight-system}, AMD's ROCprof~\cite{rocprof} and OmniTrace~\cite{omnitrace}, and Intel's VTune~\cite{WinNT} and Profiling Tools Interfaces for GPU (PTI-GPU)~\cite{pti-gpu}  provide a tracing and profiling frameworks targeting their respective GPUs. While these tools provide mechanisms to trace and profile GPU-accelerated applications, however, their capabilities are largely confined to programming models designed for their respective hardware platforms and lack portability.

Several open-source tools have been developed or extended to support heterogeneous system architectures. Established tools such as TAU \cite{tau},  HPCToolkit \cite{hpctoolkit} and Score-P \cite{scorep} has been in the HPC space for long time. TAU is a portable profiling and tracing toolkit used for performance measurement and analysis. In recent years, it has been extended to support heterogeneous systems, enabling it to monitor GPU activities through vendor provided interfaces like CUTPI, ROCm and Level-Zero. It supports instrumentation and sampling-based performance data gathering and utilizes analysis tools to generate aggregate profiles, and event traces in the form of timeline.  Similarly, HPCToolkit~\cite{hpctoolkit}, is a performance measurement and analysis tool for heterogeneous systems. It collects call path profiles and traces of applications for performance analysis and also gathers hardware counters though {\tt perf} events.  Tools like Score-P \cite{scorep}, and Caliper \cite{caliper} also offer both profiling and sampling-based performance analysis. However, THAPI stands out with unique and complementary capabilities, demonstrated by the nature of the information it gathers, the generation of tracepoints, and its comprehensive support for major heterogeneous programming models.  Notably, Lttng-hsa \cite{opencl1} and  Lttng clust\cite{opencl2} are the closest tools to ours, both built on top of LTTng.  However, these tools focus on specific subsets of programming models, with Lttng clust \cite{opencl2} targeting the OpenCL programming environment and Lttng-hsa \cite{opencl1} focusing on the ROCr runtime.  

\section{THAPI}\label{sec:thapi}

Tracing is a well-known performance analysis technique that captures the sequence of runtime events and their timing during program execution. The events are collected using tracepoints placed through instrumentation at required points in the code statically or dynamically. THAPI is a heterogeneous API tracing and profiling tool that collects host and device runtime behavior.  It utilizes the Linux Trace Toolkit Next Generation (LTTng) \cite{Desnoyers2006TheLT} for event tracing and offers complementary analysis tools, based on Babeltrace2 library, to analyze traces and provide actionable insights. THAPI utilizes Perfetto~\cite{perfetto} for timeline visualization.  In this section, we first introduce LTTng, the tracing framework integrated into THAPI’s development. We then provide a high-level overview of THAPI, followed by a detailed discussion of tracepoints and analysis tools generation methodology within the framework.

\subsection{LTTng}
The event tracing is facilitated through the utilization of the low-level tracing framework: Linux Tracing Toolkit Next Generation (LTTng). LTTng is an open-source, state-of-the-art tracing infrastructure for Linux systems. It supports kernel-space and userspace tracing via LTTng-modules and  LTTng-UST, respectively. It is a well maintained and established infrastructure used in leading data-centers. It utilizes lockless, per-CPU ring buffers for both the kernel and userspace tracing, avoiding inter-core communication and achieving low overhead and high throughput. If the application produces more events than can be consumed by the disk, LTTng drops these events rather than blocking the execution. 

LTTng's traces have a Common Trace Format (CTF), a standardized binary format optimized for performance. The traces can be parsed with a Babeltrace2 tool into a human-readable text format. 
With a tracepoint overhead in the order of nanoseconds \cite{Fournier2010CombinedTO} and the capability of its relay daemons to stream over the network, LTTng stands as an ideal solution for deployment on a large scale. We chose LTTng for THAPI because of its efficiency, performance-focused design, compatibility with various Linux systems, and availability of trace processing and analysis tools, which makes it an ideal tracing tool for our requirements. 

\subsection{Overview}

\begin{figure*}[!h]
\vspace{-20pt}
    \centering
    \begin{subfigure}[b]{0.35\textwidth}
        \includegraphics[width=\textwidth]{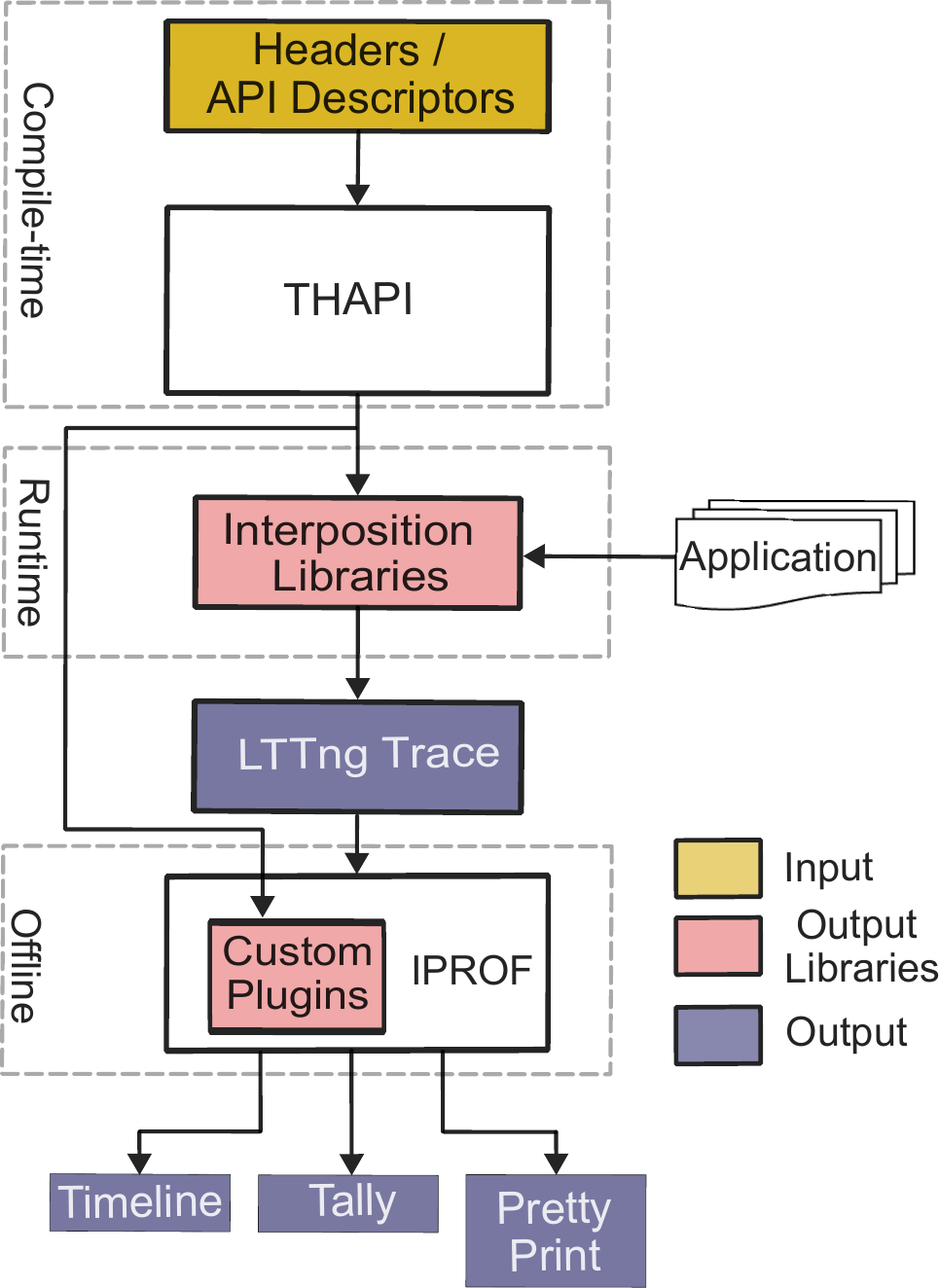}
        \caption{THAPI’s workflow for analysis of HPC applications}
        \label{fig:overall1}
    \end{subfigure}
    \hspace{20pt} % Adjust this spacing to control the gap
    \begin{subfigure}[b]{0.4\textwidth}
        \includegraphics[width=\textwidth]{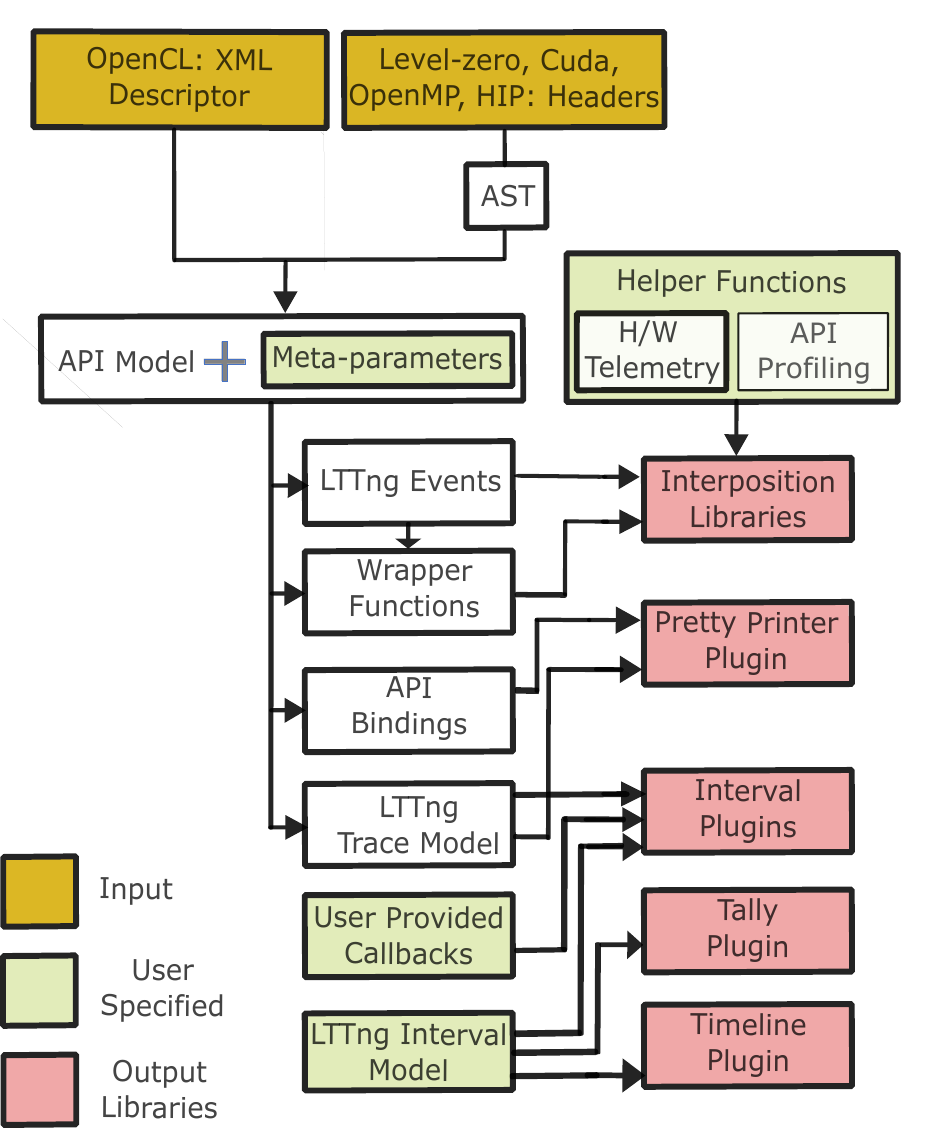}
        \caption{Internal Architecture of THAPI}
        \label{fig:Thapi}
    \end{subfigure}
    \caption{Overview of THAPI}
    \label{fig:main}
\end{figure*}
\iffalse
\begin{figure*}[!h]
    \centering
    \begin{subfigure}[b]{0.35\textwidth}
        \includegraphics[width=\textwidth]{FIG/highlevel.pdf}
       \caption {THAPI’s workflow for analysis of HPC applications }  \label{fig:overall1}
        \end{subfigure}
    \hfill
    \begin{subfigure}[b]{0.45\textwidth}
        \includegraphics[width=\textwidth]{FIG/thapi_new1.pdf}
        \caption {Internal Architecture of THAPI } 
        \label{fig:Thapi}
        \end{subfigure}
    \caption{Automatic Tracepoint Generation}
    \label{fig:main}
\end{figure*}
\fi
%TODO Say something about the picture
As we can see from  Figure~\ref{fig:overall1}, THAPI can be viewed as a tool comprising two logically distinct components: one for trace collection and another for trace analysis. The trace collection component performs programming-model centric tracing through interception library. It traces all API entry and exit points (OpenCL, CUDA, Level-Zero, HIP, MPI) or tracing callbacks (OMPT),  preserving low-level details, arguments and results from each point. The trace model is automatically generated, derived from headers or API XML descriptions (in the case of OpenCL), as illustrated in Figure~\ref{fig:overall1}.  The trace parsing utilizes the Babeltrace2 library (IPROF) and custom plugin tools generated automatically from the API model, producing various views of the trace, including Pretty Print (text), Tally (summary), and Timeline (visualization). 

 Tracing presents significant challenges, especially when we save everything in relation to the API calls. These challenges include runtime overhead, managing large data volumes, and balancing granularity with performance impact. THAPI addresses these issues through two main strategies. First, it employs selective event tracing, enabling the activation or deactivation of specific events for tracing \cite{Desnoyers2006TheLT}. It also offers the ability to selectively trace specific groups of ranks in a large-scale setting.  Second, it performs offline analysis of the collected traces effectively reducing runtime overhead. 
\iffalse
In addition to tracing APIs, THAPI has the functionality of sampling hardware telemetries such as power consumption, operating frequency, memory statistics, and compute utilization. This characteristic is essential for identifying inefficiencies.
\fi
%Todo; This part will be extended in the paper

\subsection{Automatic Tracepoint Generation}

Tracepoints are hooks inserted into a code to enable tracing of specific events.  As the complexity and diversity of programming models and APIs continue to grow, manual management of tracepoints becomes difficult. To tackle the challenges posed by the growing number of tracepoints, the THAPI tracing framework relies on the automatic generation of tracepoints. We use a systematic approach that harnesses automation and structured data extraction to ensure that all relevant events are traced comprehensively.
Figure~\ref{fig:Thapi} illustrates the complete process of automatic tracepoint and analysis plugin tools generation. In this section, we focus specifically on the interception library and the rich tracepoint generation process. 
The process begins by parsing  the API headers or description, depending on the specific programming model utilized. For CUDA, Level-Zero, OpenMP, and HIP, headers are parsed to capture details about the APIs. For OpenCL, the structured data is accessed directly from the XML API description. This information is parsed in to intermediary YAML file, that we call the {\tt API model}. From the {\tt API model} we can directly generate the interception library and tracepoints. However, this approach only gives access to the arguments on the stack, lacks detailed information (E.g. input or output memory content, structures passed by reference, etc...). Moreover, GPU timing information is not accessible, as shown in Scenario 1, Figure~\ref{fig:tracing_comparison}. For example, whether a pointer is in or out, and the value behind the pointer argument cannot be inferred directly from the headers alone, necessitating the inclusion of this expert knowledge as supplementary metadata. 
\iffalse
\begin{figure}[h]
\vspace{-20pt}
\centering
\caption{Comparison of Fully Automatic and Hybrid Tracing Approaches}
\label{fig:tracing_comparison}

\begin{tiny}
\noindent
\begin{minipage}{0.48\linewidth}
\vspace{3pt}
\begin{lstlisting}[language={},frame=none]

        Scenario 1: Fully Automatic Approach
        
              
(FULLY AUTOMATIC)
Headers -------------> Interception-Library,
                         + Tracepoints

Missing:
  - Semantic Metadata 
  - GPU Timing

\end{lstlisting}
\end{minipage}%
\hfill
\begin{minipage}{0.5\linewidth}

\vspace{10pt}
\begin{lstlisting}[language={},frame=none]
                     Scenario 2: Hybrid Approach
                     
                     
(FULLY AUTOMATIC)
Headers -------------------------> Interception-Library,
                             |           + Rich-Tracepoints 
HAND-WRITTEN)                |
Semantic Metadata --------|
(in pointer, out pointer) |
                             |
GPU Profiling Code -------|
(with stream and event profiling,
 event record start/stop)
\end{lstlisting}
\end{minipage}
\end{tiny}
\vspace{-20pt}
\end{figure}

\fi

 \begin{figure}[tbh!]
 \vspace{-20pt}
\begin{center}
\includegraphics[width=0.6\columnwidth]{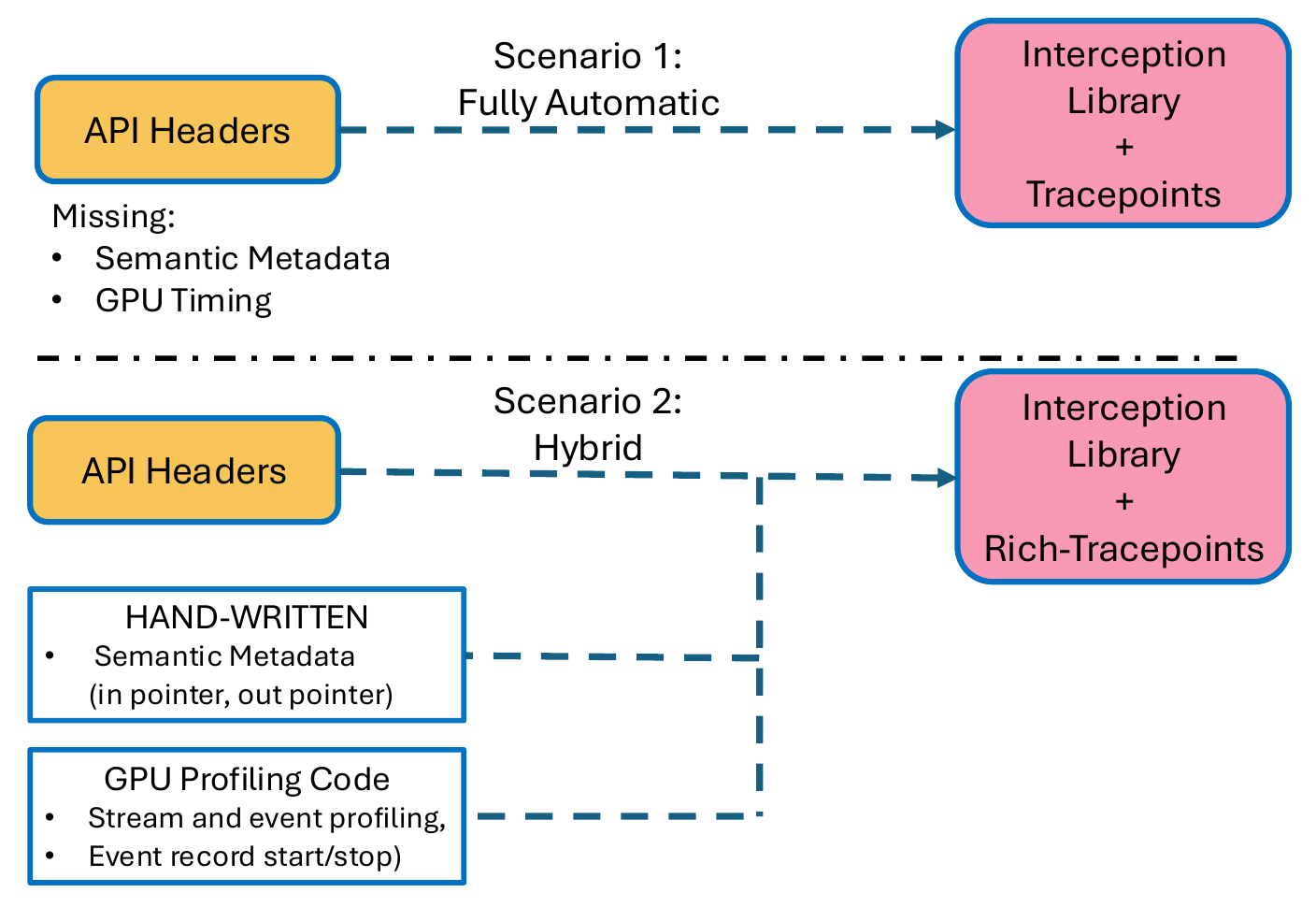}

\caption {Comparison of Fully Automatic and Hybrid Tracing Approaches}\label{fig:tracing_comparison} 
\vspace{-20pt}
\end{center}
\vspace{-20pt}
\end{figure}

\begin{figure}[tbh!]

\begin{center}
\includegraphics[width=0.7\columnwidth]{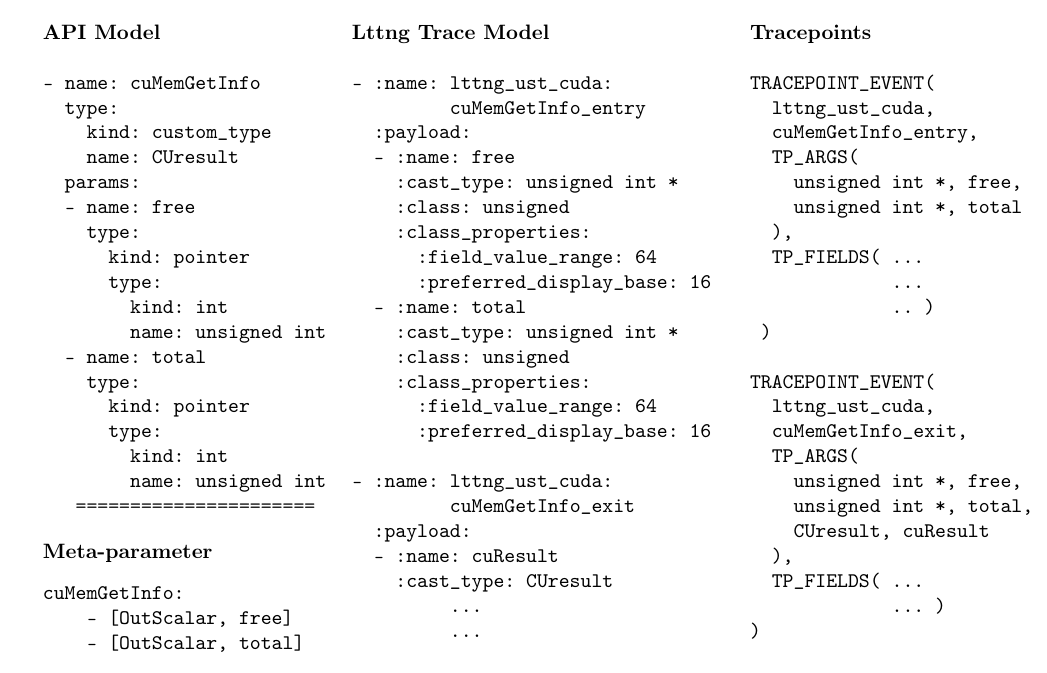}
\caption {Tracepoint generation steps for {\tt cuMemGetInfo}} \label{fig:API_model}
\end{center}
\vspace{-25pt}
\end{figure}

% Example of referencing the figure
\bigskip
In Scenario 2, the API model is enriched by user provided semantic metadata (in pointer / out pointer) and GPU Profiling Code to capture GPU timings (Cuda record entry / record exit, before submission, Level-Zero profiling / get the info during wait...) and generates the interception library and tracepoints.  THAPI utilizes this approach where the {\tt API model}  combined with user-provided {\tt Meta-Parameters}, transforming into rich {\tt LTTng Events}, and {\tt Wrapper Functions},  which provide a seamless integration point for tracing within the application code,  thereby providing a streamlined methodology to create comprehensive, user-customized model for event tracing as shown in Figure~\ref{fig:Thapi}. The {\tt Helper Functions} implement the  GPU profiling code that  captures GPU timings, find kernel details and  monitor device telemetry.  The {\tt LTTng Trace Model}, derived from the {\tt API model}  is essential for the generation of the Babeltrace2 based plugin tools: {\tt Pretty print plugin and Interval plugins}. {\tt Interval plugins} enable detailed timing analysis based on the start and end times of events. Figure~\ref{fig:API_model} illustrates the translation of {\tt API Model} to {\tt Trace Model} and {LTTng events} for the {\tt cuMemGetInfo} API call. This structured, multi-phase procedure ensures accurate and efficient translation of API descriptions into actionable tracepoints.  \\
%Todo The helper function used for find kerne details 
\iffalse
From this interception library, we can automatically generate wrapper functions and the trace model that encapsulate the tracing logic.  The trace model defines the structure and behavior of the tracing process, ensuring that all required data is captured accurately.

Finally, the tracepoints are automatically generated from the trace model. These tracepoints are strategically placed within the codebase to capture specific events or actions during API interactions. As a result, our framework enables comprehensive and automatic tracing of all relevant API activities, ensuring that nothing is overlooked. 

\fi
\iffalse
After compilation into an interposition library, tracing is enabled by loading the libraries. THAPI can be launched by the wrapper scripts -- {\tt tracer\_opencl.sh}, {\tt tracer\_ze.sh}, {\tt tracer\_cuda.sh} -- targeting individual programming models: OpenCL, Level-Zero, and CUDA respectively or  using  {\tt iprof}, a wrapper around all the tracers. The command-line usage of {\tt iprof} is as follows:
\newline

{\tt   ./iprof \-- [options] \-- ./application} \\
\fi
\iffalse
\newline
\begin{figure}[tbh!]
\begin{center}
\includegraphics[width=0.5\columnwidth]{options.png}
\caption {THAPI commandline options} \label{fig:cost}
\end{center}
\end{figure} 
\fi
%The commandline options allow filtering events, choosing tracing modes, turning on or off features such as hardware telemetry, and specifying parsing and analysis types for the collected traces according to the user's needs.
In summary, THAPI relies on automation to generate tracepoints automatically due to the difficulty of manual management. This automated approach ensures thorough and consistent tracing across different programming models and APIs. It also makes THAPI easy to maintain,  as it only requires updating the meta-parameters for the few added functions when one of the supported programming models is updated.  In theory, other tools also can re-use our LTTng tracepoints for their own applications or tools.
 %We extract information from headers or API descriptions, transform it into an intermediate YAML format, combine it with user metadata and optional tracepoints, and use this data to generate an interception library, wrapper functions, trace models, and ultimately, the tracepoints themselves.

\subsection{Babeltrace2 Analysis Tools}
The LTTng trace, once gathered, undergoes parsing and analysis using the Babeltrace2 library-based plugins tailor-made to produce specialized outputs. Babeltrace2 ~\cite{babeltrace2} is a reference parser implementation for CTF, offering a modular plugin model infrastructure that allows users to create custom plugins. We generated several plugins, incorporating source, filter, and sink components, to analyze the trace data efficiently. As we discussed in the last section, in order to overcome the time-consuming, complicated, and prone to errors nature of manually building plugins, we automated the plugin generation process. 

 To achieve this, we developed a mechanism called {\tt Metababel}, which attaches user-defined callbacks to trace events (generated automatically from the LTTng trace model). Therefore, all the plugins (implemented in C/C++ ) are collections of callbacks that are executed when they receive events. 

 \begin{figure}[tbh!]
 \vspace{-20pt}
\begin{center}
\includegraphics[width=0.4\columnwidth]{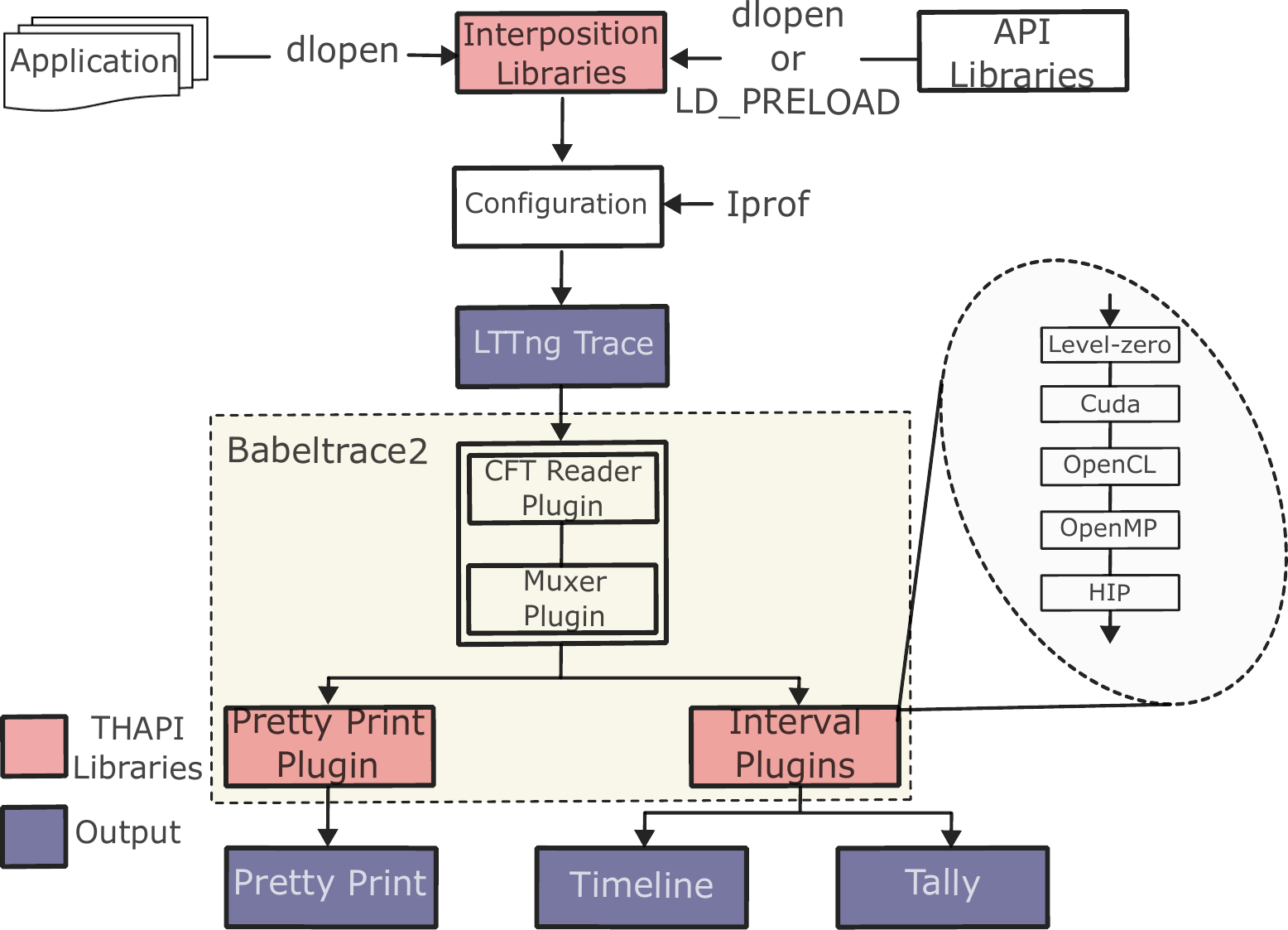}

\caption {Babeltrace2-based trace analysis tools}\label{fig:Iprof_new} 
\end{center}
\vspace{-20pt}
\end{figure} 

{\tt Metababel} abstracts Babeltrace2 details, such as reading the CTF format, unpacking fields, and generating downstream messages, simplifying post-processing scripts. This simplification streamlines post-processing scripts, allowing users to leverage its functionality without needing to understand the intricacies of Babeltrace2.    In the future, {\tt Metababel} can expand to support reading OTF traces or enhance CTF reading performance by incorporating a custom reader to bypass Babeltrace2. 

We provide these complementary plugins for generating outputs such as Pretty Print (readable text), Tally (summarized data), and Timeline (visualization), enabling diverse and comprehensive perspectives on traced data. This approach supports detailed analysis while offering flexibility in output generation to suit the specific needs and contexts of various use cases and investigations. Additionally, users can develop and use custom plugin tools to analyze the LTTng traces according to their unique requirements. Figure~\ref{fig:Iprof_new} illustrates the overall trace collection and analysis process.  Tracing begins by launching the application using the {\tt iprof} launcher, as shown below: 
\begin{center}
\texttt{./iprof {--} [options] {--} ./application}
\end{center}
 {\tt iprof}  allows filtering events, choosing tracing modes, turning on or off features such as hardware telemetry, and specifying parsing and analysis types for the collected traces according to the user's needs. Once the LTTng Traces collected, the Babeltrace2 library reads the CFT reader and Muxer plugin for serializing messages by time, and then he custom plugins then generate the desired views of the trace.

\iffalse
\begin{figure}[tbh!]
\centering
\begin{center}
\includegraphics[width=\columnwidth]{FIG/tally.pdf}
\caption {Summary of the 505.lbm\_t benchmark run on a single node of Aurora} \label{fig:timeline}
\end{center}
\end{figure} 

\fi
\iffalse
\begin{figure*}[!t]
    \centering
    \begin{subfigure}[b]{0.45\textwidth}
        \includegraphics[width=\textwidth]{FIG/tally.pdf}
        \caption{Aurora -- MPI+OMP with Level-Zero backend}
        \label{fig:sub1}
    \end{subfigure}
    \hfill
    \begin{subfigure}[b]{0.45\textwidth}
        \includegraphics[width=\textwidth]{}
        \caption{Polaris -- MPI+OMP with CUDA backend}
        \label{fig:sub2}
    \end{subfigure}
    \caption{Tally of the 505.lbm\_t benchmark run on a single node of Aurora and Polaris}
    \label{fig:main}
\end{figure*}
\fi

\subsection{Device sampling with THAPI}

Sampling device telemetry, in conjunction with API traces, provides a more holistic view of system performance and behavior. The meticulous collection of these metrics is crucial  for conducting in depth analyses on performance-to-power ratio, thermal management, and hardware-software co-optimization. In this section, we showcase the device telemetry daemon infrastructure, implemented via the Level-Zero APIs, in a concise manner. 
The Level-Zero Application Programming Interface (API) offers direct-to-metal interfaces for offloading to accelerator devices and is designed to be compatible across various compute device architectures, including GPUs, FPGAs, and other accelerator architectures. Our framework leverages Level-Zero Core and Sysman (System Management) APIs. The Core APIs are employed to initialize Level-Zero and discover drivers and available devices. Subsequently, the Sysman APIs are utilized to sample the energy, operating frequency, memory stats, fabric stats, and device utilization.

The device sampling framework is implemented as a daemon program that can be optionally enabled with THAPI using the {\tt --sample} option. When activated, it begins sampling device counters at a user-defined sampling interval, with a default period of 50ms. The collected metrics are then streamed into the LTTng trace for analysis.

\subsection{Timeline}
\vspace{-5pt}
Timeline visualization is important for identifying performance bottlenecks and enabling optimization. To visualize the traces, we utilized Perfetto~\cite{perfetto} -- a trace and visualization framework by Google. Perfetto uses a protobuf format for visualization, so we implemented a mechanism to convert the trace data into this format. The timeline is structured with multiple rows illustrating the gathered API traces and corresponding device samplings. Each compute node utilizes the topmost rows to represent the host  and device API calls respectively. As can be seen from Figure~\ref{fig:timeline}, the first row depicts the host process and the second row shows the device.  For each GPU within a node, as illustrated in Figure~\ref{fig:timeline}, there are multiple rows representing the device telemetry.

\begin{figure}[tbh!]
\vspace{-15pt}
\begin{center}
\includegraphics[width=0.5\columnwidth]{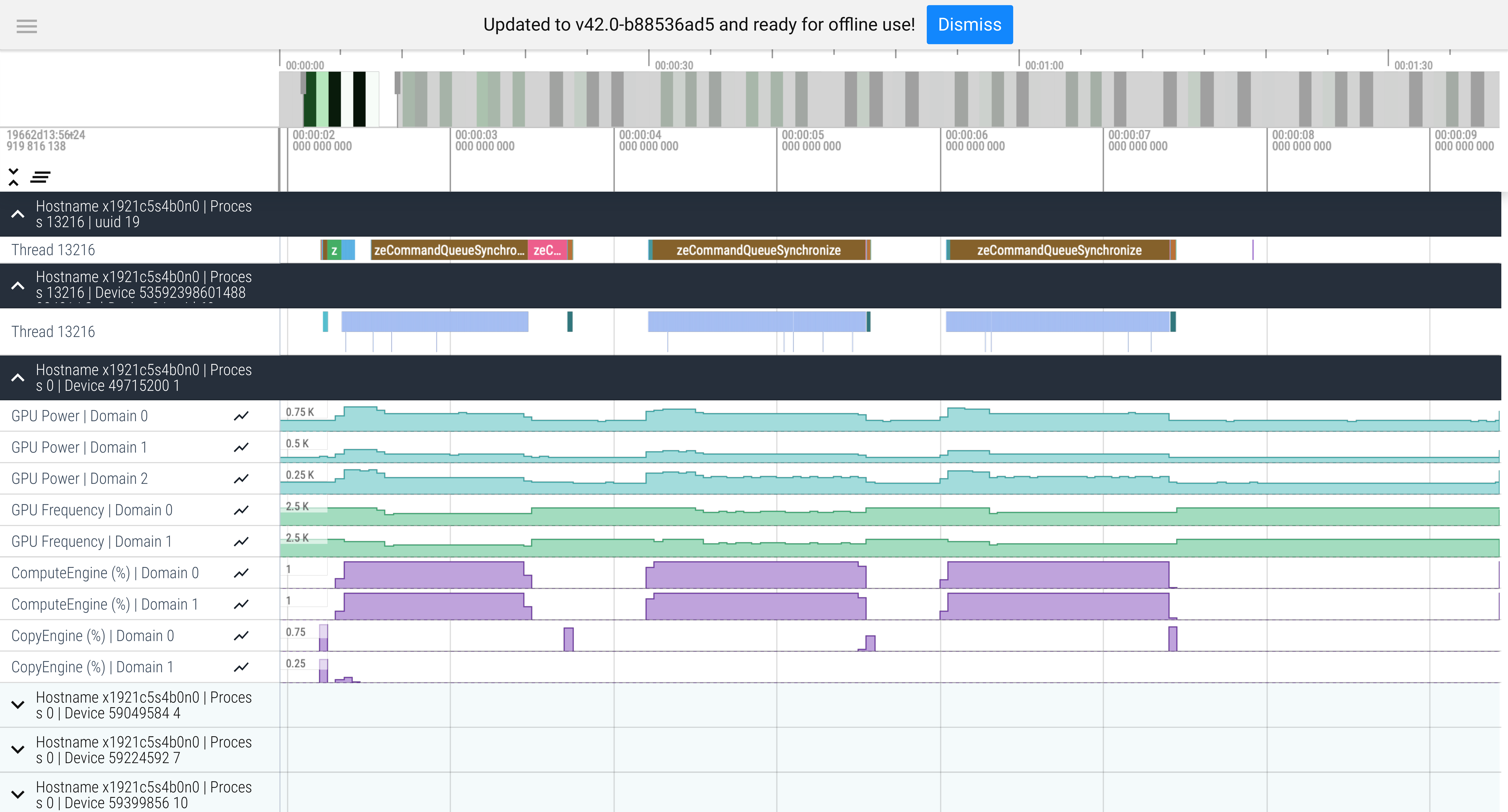}
\caption {Timeline of traces and device telemetry collected from convolution1D benchmark run on Aurora.} \label{fig:timeline}
\end{center}
\vspace{-25pt}
\end{figure} 
The first three rows are allocated for showcasing power traces from different parts of the GPU chip, specifically labeled as {\tt Power|Domain 0}, {\tt Power|Domain 1 } and {\tt Power|Domain 2}.  The first indicates the overall power usage of the chip, while {\tt Power|Domain 1} and {\tt Power|Domain 2} display the power consumption attributed to each of the two available tiles in the PVC. Subsequently, the next two rows display the operational frequencies of each tile, denoted as {\tt GPU Frequency|Domain 0} and {\tt GPU Frequency|Domain 1}. Following these, two rows are dedicated to demonstrating the utilization of the compute engine in tile 0 and tile 1, respectively. The final two rows are reserved for depicting the utilization of the copy engines in the first and second tiles. Based on the user's sampling configuration, the content of the timeline can vary.

\vspace{-10pt}
\subsection{On-node Processing}
\vspace{-5pt}
Users can choose to save only the aggregate of the trace, which is lightweight, typically in the range of kilobytes, depending on storage availability and usage requirements. These aggregates can be replayed to generate tally profiles and are the default setting for multi-node experiments. In such scenarios, each local master sends its aggregate to the global master, where the summaries are combined into a composite profile. Traces are temporarily stored in local scratchpad memory to generate these aggregates. For detailed postmortem analysis, users can enable the {\tt --trace} option to permanently save the LTTng trace for one or more specific ranks. We have experimented this on a production machine and successfully scaled up to 512 nodes run.
\vspace{-5pt}
\section{Case Studies} \label{sec:case}
\vspace{-5pt}
In this section, we present examples demonstrating the effectiveness of THAPI. These case studies highlight the unique capabilities of THAPI and its applicability to various use cases.
\subsection{Debugging OpenMP Runtime}
We utilized trace analysis to diagnose a performance issue within the OpenMP runtime, specifically  related to its use of the Copy Engine in the Level-Zero backend. Since the Intel OpenMP runtime is closed-source, direct inspection was not possible. However, by tracing Level-Zero API calls, we were able to analyze its behavior.

Our analysis revealed that the runtime did not leverage Level-Zero's capability to use a dedicated Copy Engine for data transfers. Instead, it consistently relied on the general compute engine, with all command lists bound to it. After identifying this inefficiency, we reported the issue, leading to its resolution. This case demonstrates that even in the absence of source code, access to API call traces provides sufficient context for runtime developers to analyze proprietary runtimes and report performance-related issues.

\subsection{Mitigating Undefined Behavior in Level-Zero}

In Level-Zero, certain API properties must be explicitly set to NULL. For example, the pNext pointer in {\tt zeDeviceGetProperties} must be initialized correctly. Failing to do so results in undefined behavior.
In C, it’s easy to overlook this requirement:
\iffalse
\begin{minted}{c}
ze_device_properties_t device_properties;
ret = zeDeviceGetProperties(global_ze_devices_handle[d], &device_properties);
\end{minted}

Here, {\tt device\_properties.pNext} may contain an uninitialized value, leading to unpredictable behavior. The correct approach is to either: 
\begin{minted}{c}
ze_device_properties_t device_properties = {0}; or device_properties.pNext = NULL;
\end{minted} 
\fi
\begin{verbatim}
ze_device_properties_t device_properties;
ret = zeDeviceGetProperties(global_ze_devices_handle[d], &device_properties);
\end{verbatim}

Here, {\tt device\_properties.pNext} may contain an uninitialized value, leading to unpredictable behavior. The correct approach is to either:

\begin{verbatim}
ze_device_properties_t device_properties = {0}; or device_properties.pNext = NULL;
\end{verbatim}
Bugs of this nature have been observed in real-world applications and have been reported and subsequently fixed.
To mitigate common low-level API mistakes—including missing NULL assignments, unhandled release events, and non-reset of command lists, we developed a post-mortem validation plugin.

\subsection{Analysis of HIPLZ Implementation on Aurora}

HIPLZ is a compiler and runtime system that enables HIP implementations to run on Intel GPU architectures via the Level-Zero backend. Although this enhances portability, it also introduces additional complexity, making it essential to understand the interactions between different programming models.\\
\iffalse
\begin{scriptsize}
\\
\textbf{BACKEND\_HIP, BACKEND\_ZE | 1 Hostname | 1 Process | 1 Thread}  
\begin{minted}[frame=single, fontsize=\small]{text}
                           Name |     Time | Time(%) |   Calls |  Average |  ...         
           hipDeviceSynchronize |    4.73s |  37.39% |      16 | 295.89ms |  ...        
         zeEventHostSynchronize |    4.68s |  36.99% | 9927772 | 471.80ns |  ...         
                      hipMemcpy |    1.77s |  13.98% |       7 | 252.79ms |  ...         
       __hipUnregisterFatBinary | 500.91ms |   3.96% |       1 | 500.91ms |  ...         
  zeCommandListAppendMemoryCopy | 394.50ms |   3.12% |       7 |  56.36ms |  ...         
                hipLaunchKernel | 262.70ms |   2.07% |      32 |   8.21ms |  ...         
                 zeModuleCreate | 256.09ms |   2.02% |       1 | 256.09ms |  ... 
   ......................................//..................................
\end{minted}
\end{scriptsize} 
\fi
\begin{scriptsize}

\vspace{2pt}

\vspace{2pt}
\centering
\begin{verbatim}
                     BACKEND_HIP,BACKEND_ZE | 1 Hostnames | 1 Processes | 1 Threads | 

                           Name |     Time | Time(%) |   Calls |  Average |      Min |      Max |         
           hipDeviceSynchronize |    4.73s |  37.39% |      16 | 295.89ms |    678ns | 867.22ms |         
         zeEventHostSynchronize |    4.68s |  36.99% | 9927772 | 471.80ns |    390ns |   3.56ms |         
                      hipMemcpy |    1.77s |  13.98% |       7 | 252.79ms | 202.40ms | 291.56ms |         
       __hipUnregisterFatBinary | 500.91ms |   3.96% |       1 | 500.91ms | 500.91ms | 500.91ms |         
  zeCommandListAppendMemoryCopy | 394.50ms |   3.12% |       7 |  56.36ms |  48.86ms |  69.42ms |         
                hipLaunchKernel | 262.70ms |   2.07% |      32 |   8.21ms |   9.71us | 261.35ms |         
                 zeModuleCreate | 256.09ms |   2.02% |       1 | 256.09ms | 256.09ms | 256.09ms |         
 
   .............................................//.................................................
  
\end{verbatim}

\end{scriptsize}

The table above is a partial snapshot of the summary of the THAPI trace for the mini-app  Local Response Normalization (LRN), which is written in HIP and executed on the Aurora (Intel architectures) using HIPLZ. This evaluation provides insights into performance characteristics of the implementation and the overhead caused by the layering. For example, we can analyze how {\tt hipDeviceSynchronize} implemented on top of {\tt zeEventHostSynchronize} spin lock and its impact on performance. 

\begin{figure}[tbh!]
\begin{center}
\includegraphics[width=0.9\columnwidth]{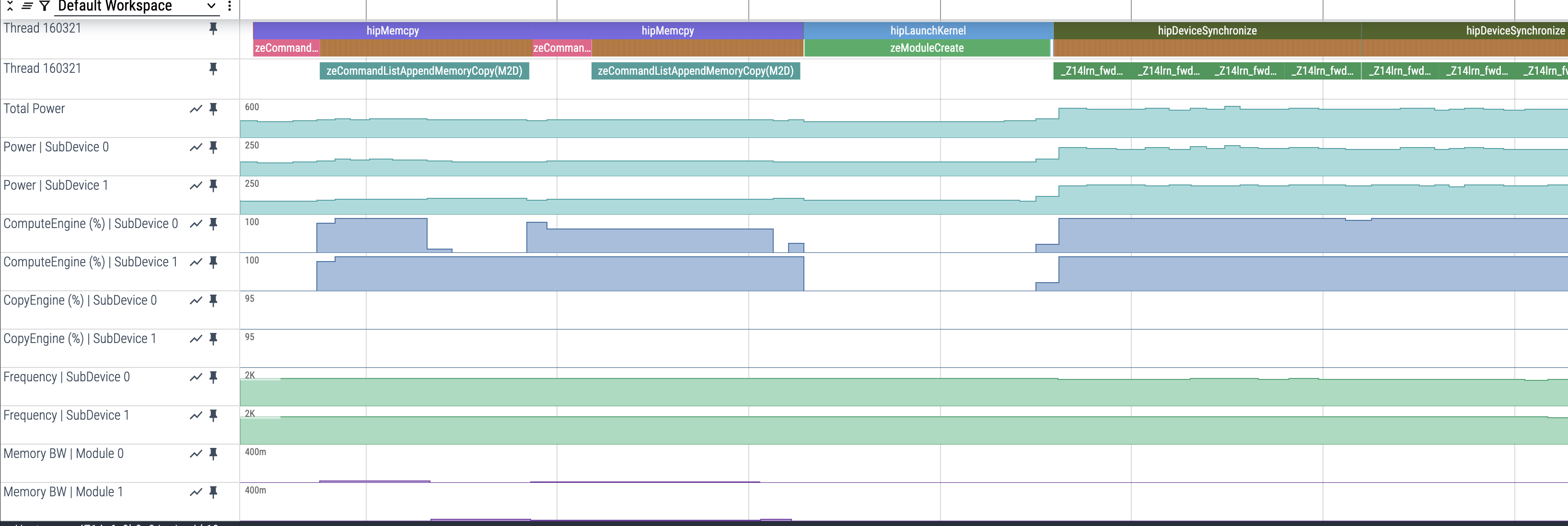}
\caption {Timeline for HIP implementation of LRN benchmark on Aurora} \label{fig:timeline_hip}
\end{center}
\vspace{-20pt}
\end{figure}

Figure~\ref{fig:timeline_hip} shows a timeline visualization of the experiment. The first row, representing host API calls, depicts the overlap between HIP and Level-Zero layers. The second row illustrates device API calls. The remaining rows display device telemetry data, offering deeper insights into runtime behavior.

 \section{Evaluation} \label{sec:eval}
This section outlines the hardware configuration utilized in our experiments and details the benchmarks employed. Furthermore, we present the results of our experiment.
\subsection{Experimental Setup}
We validated THAPI on two HPC systems at the Argonne Leadership Computing Facility: Aurora and Polaris. Aurora, built on the HPE Cray-Ex platform, consists of 10,624 nodes, while Polaris, an HPE Apollo 6500 Gen 10+ system, features 560 nodes. The node configurations for both systems are detailed in Table~\ref{tab:system_config}. 

\begin{table}[h]
\vspace{-20pt}
\centering
\small
\caption{System Configurations}
\label{tab:system_config}
\begin{tabular}{|l|l|l|}
\hline
\textbf{Component} & \textbf{Aurora} & \textbf{Polaris} \\
\hline
CPU per Node  & Intel Xeon CPU Max 9470C  & AMD EPYC Milan 7543P  \\ 
Cores/Threads per CPU  & 52/104   & 32/64   \\ 
GPU  & Intel Data Center Max 1550  & NVIDIA A100  \\ 
GPUs per Node  & 6  & 4  \\ 
No. of Nodes  & 10,624  & 560  \\ 
Programming Model Backend & Level-Zero & CUDA\\
\hline
\end{tabular}
\vspace{-20pt}
\end{table}

\iffalse
\begin{figure}[tbh!]
\centering
\begin{center}
\includegraphics[width=0.4\columnwidth]{FIG/aurora.pdf}
\caption {Aurora node architecture \cite{aurora} } \label{fig:aurora}
\end{center}
\end{figure} 
\fi
\iffalse
The Xeon CPUs each have 52 cores, capable of handling two hardware threads per core, and are outfitted with 64GB of HBM. The PVC is built on the Xe Core architecture. Each Xe core is composed of 8 vector and 8 matrix engines, supported by 512 KB of L1 cache, configurable as cache or Shared Local Memory (SLM). They are interconnected using the Intel XeLink interfaces. Every node includes 8 HPE Slingshot-11 Network Interface Cards (NICs), and the entire system is structured in a dragonfly network topology. A group of 16 Xe cores forms a slice, and 4 such slices are combined with a substantial L2 cache and 4 HBM2E memory controllers to create a stack or tile. 

\subsubsection{Polaris}
Polaris is an HPE Apollo 6500 Gen 10+ system featuring 560 nodes. Each node is powered by a single 2.8 GHz AMD EPYC Milan 7543P with 32-core CPU, 512 GB of DDR4 RAM, and four NVIDIA A100 GPUs connected via NVLink. Additionally, each node comprises two local 1.6TB SSDs in RAID0 and two Slingshot 11 network adapters. The 560 nodes are organized with two nodes per chassis, seven chassis per rack, and 40 racks.
\fi
\subsubsection {Benchmarks}
We used HeCBench~\cite{hecbench} and SPEChpc 2021~\cite{spechpc2021} benchmark suites for the experiments. The HeCBench, short for Heterogeneous Computing Benchmark Suite, is an assemblage of various samples, benchmarks, and mini-applications derived from numerous open-source projects.  Given that the device telemetry sampling interval is set at 50ms, we opted for benchmarks that run for a minimum of five seconds. Consequently, we selected a total of 70 benchmarks from the suite. In addition to HeCBench, we also tested THAPI using  the MPI+OMP target offload version of the SPEChpc 2021 benchmark suite.  We utilized the SPEChpc benchmarks on both Aurora and Polaris. 
\iffalse In addition, we conducted experiments on HIP codes, which are not included here due to incomplete time and space analysis. In these experiments, we successfully traced HIP programs on Aurora using the HIPLZ implementation—a compiler and runtime system that leverages the Intel Level-Zero API to support HIP.
\fi

\subsection{Experiment and Results}

We executed the benchmarks with THAPI across various tracing modes: minimal, default, and full, each distinguished by the quantity of events THAPI tracks providing trade-off between space requirement and detail. Below is a concise definition of each tracing mode:
\begin{itemize}

\item Minimal: Captures kernel execution events, including timings, names, and device commands.
\item Default: Captures all events except non-spawned APIs (e.g., cuQueryEvent, mpiEventReady) invoked in spin-lock scenarios.
\item Full: Captures all events without exclusions, intended exclusively for debugging purposes.
\end{itemize}

For each setting, we performed the experiments both with and without device sampling (telemetry), documenting the performance overhead for each of the six configurations relative to the baseline run of the benchmark. We call the three benchmark runs without sampling as {\tt T-min}, {\tt T-default} , {\tt T-full} and the other three runs with sampling as {\tt TS-min}, {\tt TS-default}, {\tt TS-full}.

\iffalse
\begin{table}
\caption{Experimental scenarios } \label{scenario}
\begin{center}
\begin{tabular}{ |l|l|l| } 
\hline
Tracing-mode & No-sampling & Sampling \\
\hline
Minimal & {\tt T-min} & {\tt TS-min} \\ 
Default & {\tt T-default} & {\tt TS-default} \\ 
Full & {\tt T-full} & {\tt TS-full} \\
\hline
\end{tabular}
\end{center}
\end{table}

\begin{figure}[tbh!]
\begin{center}
\includegraphics[width=0.5\columnwidth]{FIG/thapi_result1.pdf}
\caption {Average performance overhead of THAPI under different configurations (Aurora node)} \label{fig:results}
\end{center}
\end{figure} 
\fi

\subsubsection{Tracing Overhead Analysis}
The illustration of the runtime overhead incurred by tracing API calls within the programming model across various tracing modes of THAPI, both with and without device sampling, is presented Figure~\ref{fig:results}. 
The {\tt T-default} demonstrates an average overhead of 5.36\% with the median at 1.99\%. This indicates that while capturing essential information necessary for reconstructing the application's state, THAPI manages to do so without imposing considerable overhead. Although in {\tt T-minimal} tracing mode THAPI monitors fewer events than {\tt T-default}, it experiences a slightly higher overhead. Nonetheless, the volume of data gathered and the time required for its processing are substantially reduced. Adding device sampling introduces an approximate average additional runtime cost of one percent compared to running THAPI without sampling.

\begin{figure*}[!t]
\vspace{-20pt}
    \centering
    \begin{subfigure}[b]{0.45\textwidth}
        \includegraphics[width=\textwidth]{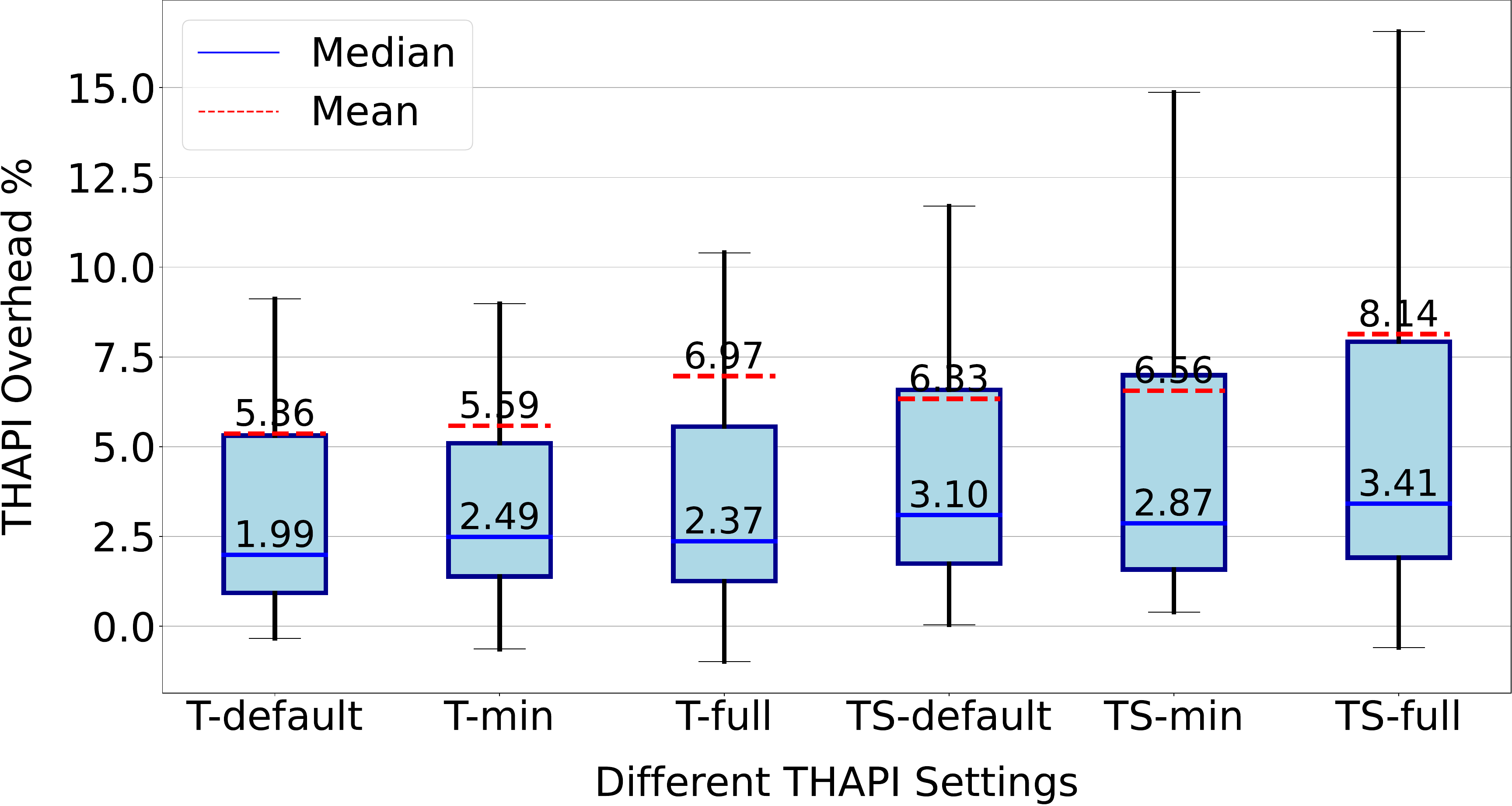}
        \caption{Average performance overhead of THAPI under different configurations (HeCBench on Aurora node)}
        \label{fig:results}
    \end{subfigure}
    \hfill
    \begin{subfigure}[b]{0.45\textwidth}
        \includegraphics[width=\textwidth]{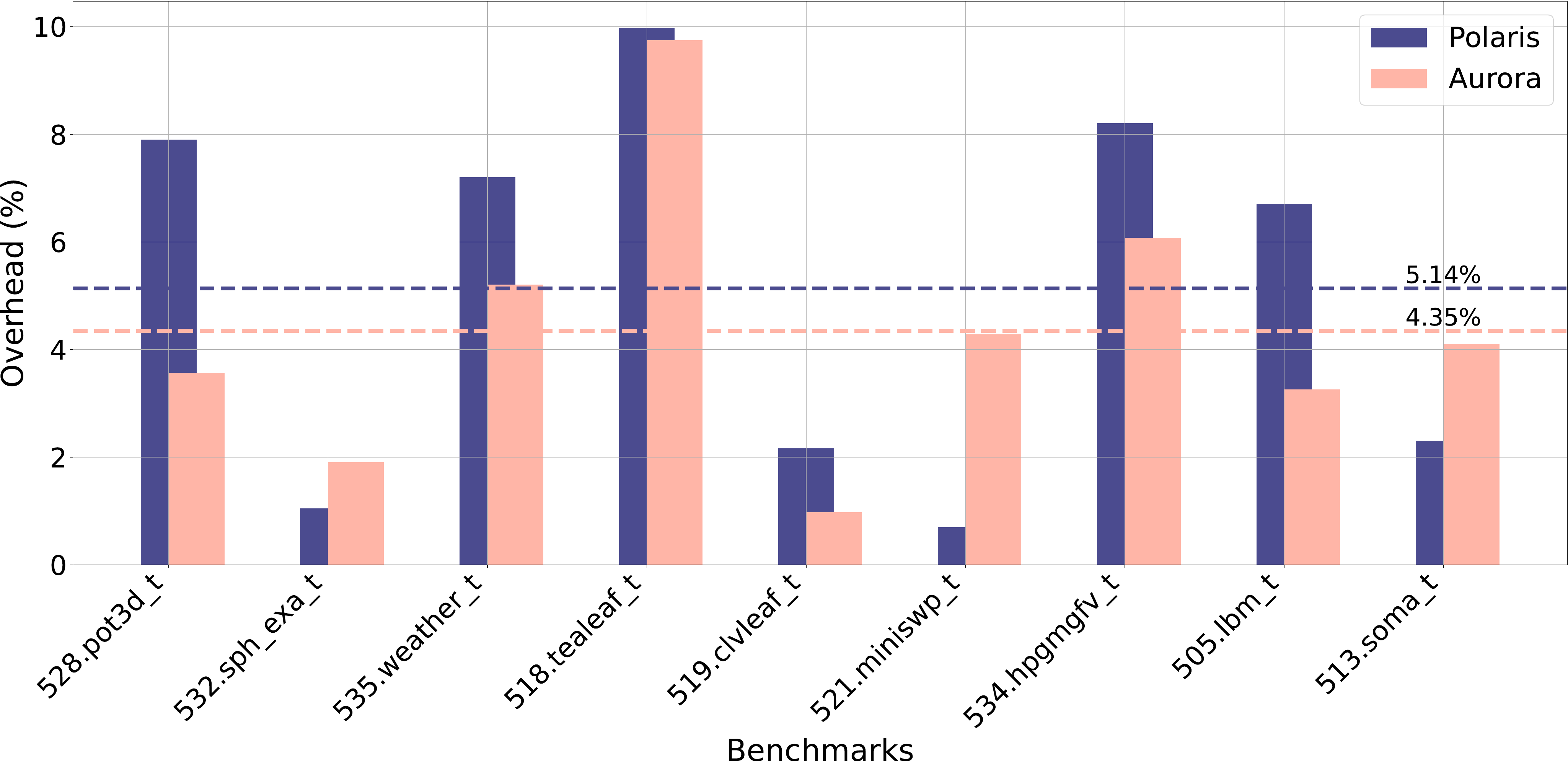}
        \caption{Percentage of runtime overhead of THAPI on Polaris vs Aurora}
        \label{fig:polaris}
    \end{subfigure}
    \caption{Performance evaluation of THAPI}  
    \label{fig:main}
    \vspace{-20pt}
\end{figure*}
\iffalse
\begin{figure}[tbh!]
\centering
\begin{minipage}{0.45\textwidth} % Adjust width as needed
    \centering
    \captionof{table}{Experimental scenarios}
    \label{scenario}
    \begin{tabular}{ |l|l|l| } 
    \hline
    \textbf{Tracing-mode} & \textbf{No-sampling} & \textbf{Sampling} \\
    \hline
    Minimal & {\tt T-min} & {\tt TS-min} \\ 
    Default & {\tt T-default} & {\tt TS-default} \\ 
    Full & {\tt T-full} & {\tt TS-full} \\
    \hline
    \end{tabular}
\end{minipage}
\hfill
\begin{minipage}{0.45\textwidth} % Adjust width as needed
    \centering
    \includegraphics[width=\linewidth]{FIG/thapi_result1.pdf}
    \caption{Average performance overhead of THAPI under different configurations (Aurora node)}
    \label{fig:results}
\end{minipage}
\end{figure}
\fi

\iffalse
\begin{figure}[tbh!]
\begin{center}
\includegraphics[width=0.8\columnwidth]{FIG/Rate_of_calls.pdf}
\caption {Rate of calls distribution for HeCBench benchmarks in default tracing mode} \label{fig:cost}
\end{center}
\end{figure} 

\fi

We also performed experiments using {\tt SPEChpc 2021}-tiny benchmark suite -- the MPI + OMP target offload version -- utilizing all the available six GPUs on Aurora node and the four GPUs on Polaris . As we can see from ~\ref{fig:polaris}, the mean  tracing overheads (default-mode) for the whole benchmark suite on Aurora is 4.35\%. The maximum overhead by a benchmark does not exceed 10\%.   
\iffalse
\begin{figure}[tbh!]
\begin{center}
\includegraphics[width=0.6\columnwidth]{FIG/SPEC1.pdf}
\caption {Run times and average overhead of THAPI on Aurora} \label{fig:results1}
\end{center}
\end{figure} 
\fi

On the other hand,  the experiments on Polaris demonstrated a mean tracing overhead of 5.14\% as shown in  Figure~\ref{fig:polaris}. We also witnessed performance variation among benchmarks when they run on Aurora and Polaris. Some applications, such as {\tt 532.sph\_exa, 521.miniswp} achieved better time to completion on Polaris while others, {\tt 505.lbm, 519.clvleaf}, performed better on Aurora.

\subsubsection{Space Requirement Assessment}
We also analyzed the space requirement for the traces. As we can see from Figure ~\ref{fig:space}, the minimal tracing modes ({\tt T-min} and {\tt TS-min} ) consistently demonstrated the lowest space requirements across all benchmarks for the SPEChpc 2021 runs on Aurora node. The full tracing modes ({\tt T-full} and {\tt TS-full} ) require significantly more space, with sampling ({\tt TS-full}) further increasing the space demand. Benchmarks such as 534.hpgmgfv\_t and 521.miniswp\_t shows the largest differences between minimal and full tracing modes, both with and without sampling.

\iffalse
\begin{figure}[tbh!]
\begin{center}
\includegraphics[width=0.5\columnwidth]{FIG/space.pdf}
\caption {Disk space requirement for SPEChpc 2021 benchmarks across different tracing-modes of THAPI} \label{fig:space}
\end{center}
\end{figure} 
\fi
As we mentioned earlier, the {\tt T-full} tracing mode is intended only for debugging purposes. The {\tt T-default} mode captures all the necessary information to recreate the context of an application run. We also analyzed the relative space requirements between the tracing-modes. We also analyzed the relative space requirements across the tracing modes. On average, the default and minimal tracing modes require less than 20\% and 17\% of the space needed by the full mode, respectively. The users have also the option to save only the summary of the trace without requiring to save the traces permanently if a high-level overview meets their requirements.
\iffalse
\begin{figure}[tbh!]
\begin{center}
\includegraphics[width=0.5\columnwidth]{FIG/comparison1.pdf}
\caption {Normalized space requirement per tracing-modes} \label{fig:iprof_results}
\end{center}
\end{figure} 
\fi

\begin{figure*}[!t]
\vspace{-20pt}
    \centering
    \begin{subfigure}[b]{0.45\textwidth}
        \includegraphics[width=\textwidth]{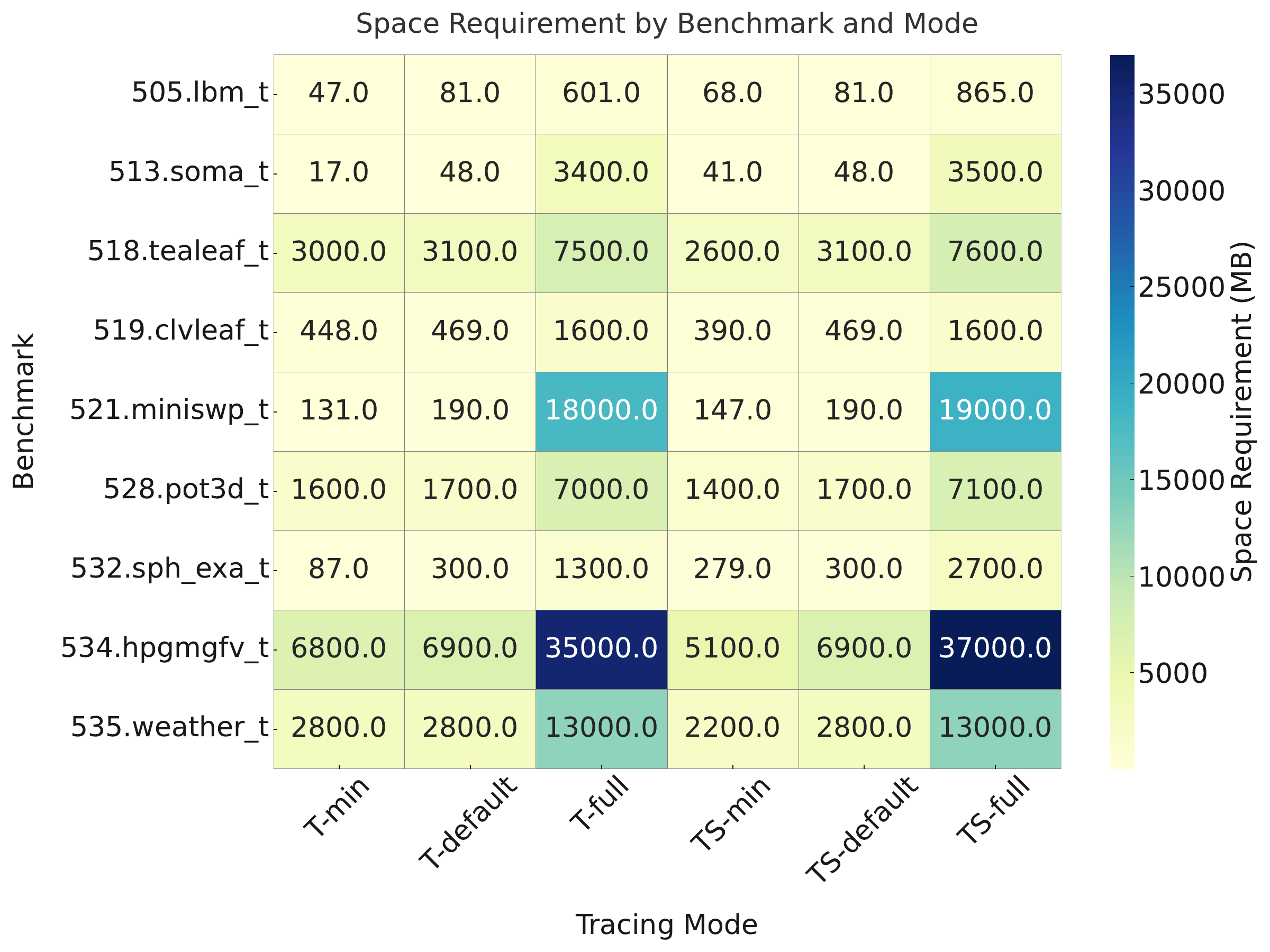}
        \caption{Disk space requirement for SPEChpc 2021 benchmarks across different tracing-modes of THAPI}
        \label{fig:space}
    \end{subfigure}
    \hfill
    \begin{subfigure}[b]{0.45\textwidth}
        \includegraphics[width=\textwidth]{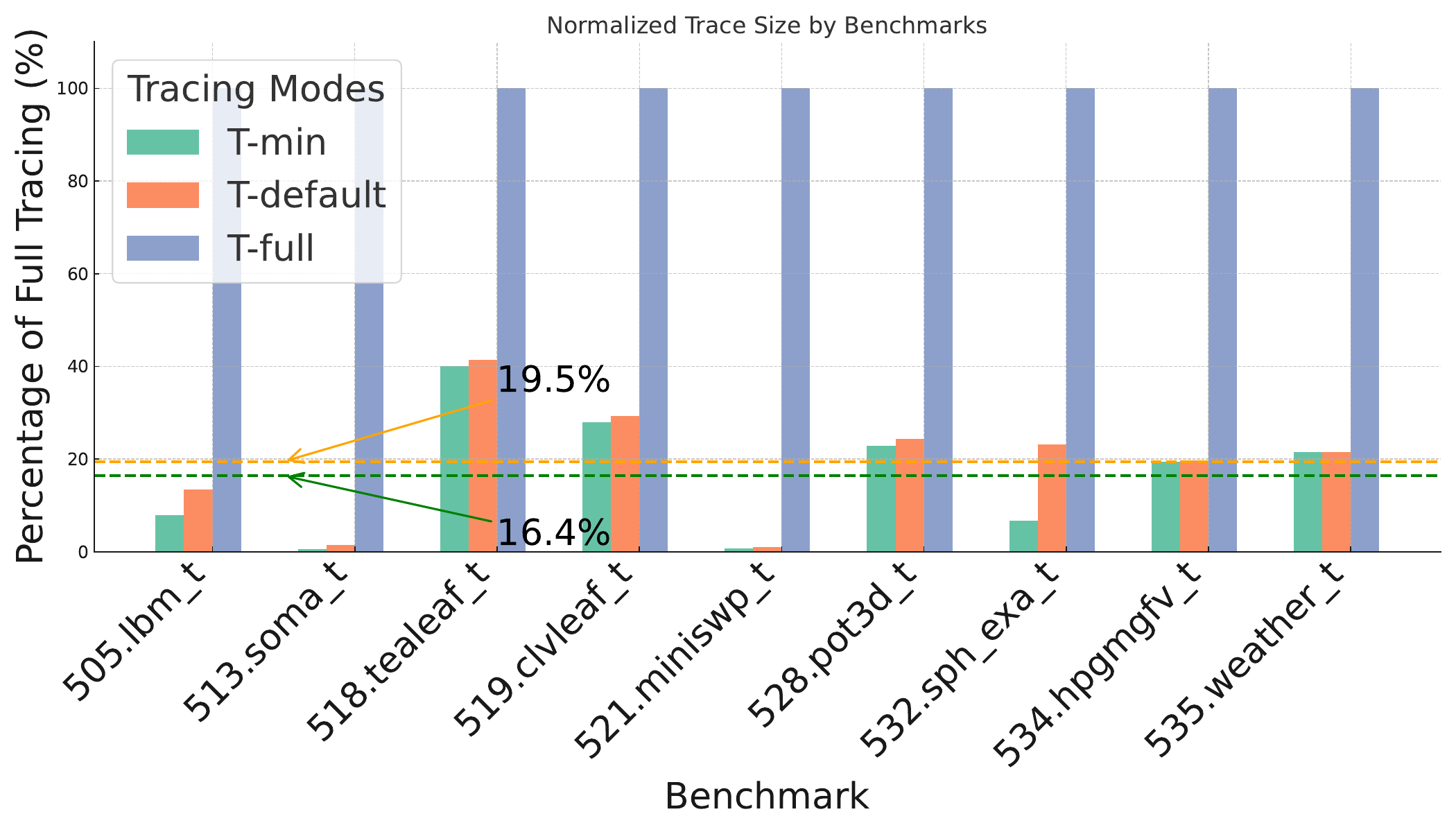}
        \caption{Normalized space requirement per tracing-modes}
        \label{fig:iprof_results}
    \end{subfigure}
    \caption{Disk space requirement of the SPEChpc 2021 traces}  
    \label{fig:main}
    \vspace{-20pt}
\end{figure*}

\section{Conclusion }\label{sec:co}
 In recent years, high-performance computing systems,  powered by  heterogeneous compute components, crossed the exascale performance barrier. Applications running on these systems navigate through different programming models to acquire their performance. 
 However, understanding application performance across diverse programming models has also became more complex requiring innovative approaches.  We proposed THAPI, a comprehensive, programming model-centric tracing tool for heterogeneous HPC systems. THAPI uses LTTng,  a very efficient Linux tracer, to collect events. In our experiments,  we validated the effectiveness of the framework through multiple case studies, and also demonstrated minimal performance overhead -- 1.99\% for HeCBench applications and 4.11\% for SPEChpc 2021 applications -- showing its efficiency.  As a future prospect, we plan to integrate machine learning technique into THAPI for advanced trace analysis and hidden pattern discovery. In addition to that, we are also working on online trace analysis, where tracing and analysis can be performed concurrently to enable adaptive optimizations during application runtime.   
 
 \section*{Acknowledgment}

 This research used resources of the Argonne Leadership Computing Facility, a U.S. Department of Energy (DOE) Office of Science user facility at Argonne National Laboratory and is based on research supported by the U.S. DOE Office of Science-Advanced Scientific Computing Research Program, under Contract No. DE-AC02-06CH11357.

\bibliography{reference}
\end{document}